\documentclass[krantz1,a4paper]{krantzmak}
\usepackage{fixltx2e,fix-cm}
\usepackage{amssymb}
\usepackage{amsmath}
\usepackage{graphicx}
\usepackage{subfigure}
\usepackage{makeidx}
\usepackage{multicol}
\usepackage{natbib}
\usepackage{graphicx}
\begin{document}
\newcommand{\lik}{\mathcal{L}}

\newcommand{\rc}{}
\newcommand{\reftitle}{}

\newcommand{\T}{^{\scriptscriptstyle\top}}

\newcommand{\dif}{\mathrm{d}}

\title{Handbook of Mixture Analysis}
\author{Gilles Celeux, Sylvia Fr\"uwirth-Schnatter, and Christian P. Robert}

\mainmatter


\setcounter{chapter}{18}

\chapter{Mixture Models in Astronomy
}    \label{chapter19}

\large{\bf Abstract}

\normalsize
\noindent 
Mixture models combine multiple components into a single probability density function. They are a natural statistical model for many situations in astronomy, such as surveys containing multiple types of objects, cluster analysis in various data spaces, and complicated distribution functions. This chapter in the CRC Handbook of Mixture Analysis is concerned with astronomical applications of mixture models for cluster analysis, classification, and semi-parametric density estimation. We present several classification examples from the literature, including identification of a new class, analysis of contaminants, and overlapping populations. In most cases, mixtures of normal (Gaussian) distributions are used, but it is sometimes necessary to use different distribution functions derived from astrophysical experience. We also address the use of mixture models for the analysis of spatial distributions of objects, like galaxies in redshift surveys or young stars in star-forming regions. In the case of galaxy clustering, mixture models may not be the optimal choice for understanding the homogeneous and isotropic structure of voids and filaments. However, we show that mixture models, using astrophysical models for star clusters, may provide a natural solution to the problem of subdividing a young stellar population into subclusters. Finally, we explore how mixture models can be used for mathematically advanced modeling of data with heteroscedastic uncertainties or missing values, providing two example algorithms, the measurement error regression model of Kelly (2007) and the Extreme Deconvolution model of Bovy et al. (2011). The challenges presented by astronomical science, aided by the public availability of catalogs from major surveys and missions, are a rich area for collaboration between statisticians and astronomers.


\section{Introduction}

Astronomy is the scientific study of objects beyond Earth: planets, stars, galaxies, and the cosmos itself.   Observations are made with ground-based and satellite-borne telescopes spanning the entire electromagnetic spectrum from radio through gamma-rays.  Data structures and scientific problems are diverse so that many statistical techniques are needed to advance  our understanding of cosmic objects and phenomena.  Mixture models have played a significant role in such analyses, though not always under this name. The method is used for many purposes, ranging from the classification of objects in a multi-dimensional parameter space to the study of spatial clustering patterns of stars or galaxies.  This second problem has attracted attention among statisticians. A galaxies dataset \citep{1986AJ.....92.1238P}, made up of recessional velocities of 83 galaxies in units of km~s$^{-1}$, has served as a challenging test case for estimating the number of components in a mixture model.

Astronomical problems involving mixture models often differ from situations familiar from social or biological sciences. For example, astronomical datasets may have unusual forms of the probability density function. Some distributions may be fit with the more conventional log-normal distributions (e.g., masses of globular clusters), Pareto distributions (e.g., initial masses of high-mass stars), or gamma distributions (e.g., galaxy luminosities). However, other, more unusual examples may originate from physical and astrophysical processes: the distributions of photon energies from an X-ray source is dictated by thermal and quantum physics; the approximate distribution of stars in a dynamically relaxed star cluster can be derived from Newtonian gravity; and the distributions of different populations of stars in a galaxy is based on the galaxy's star-formation history.

Other problems apply normal mixture models to ``Big Data'' produced by wide-field surveys.  Here, a telescope can produce exabytes of images from repeated scans of the sky, from which catalogs of billions of sources (rows) with tens of measured properties (columns) are generated.  Dozens of diverse cosmic populations may be present in the survey.  The most famous of such surveys has been the Sloan Digital Sky Survey \citep[SDSS;][]{2003AJ....126.2081A}, which initially provided spectra for several million stars and galaxies and photometry (brightness measurements) for about 500 million objects in five filters. The scale of surveys continues to grow with the planned Large Synoptic Survey Telescope  \citep[LSST;][]{2008arXiv0805.2366I} intended to monitor more than 30 billion objects over a thousand epochs during a ten-year timeframe.

Our presentation of astronomical uses of mixture models here is not systematic or comprehensive, but is designed to give a sense of the scope and challenges arising in a variety of settings.  We hope this review and commentary will encourage statisticians to share their expertise with astronomers, advancing the characterization and understanding of many facets of the Universe around us.

\section{Clusters of stars and galaxies}

Statistical methods have been important for modeling the spatial distributions of astronomical objects that are physically associated. Examples include star clusters and galaxy clusters, both of which are held together by gravity but may exhibit anisotropic and intertwined structures inherited from a complicated formation process. A variety of statistical methods have been used to examine these systems, including hierarchical clustering methods for identifying individual clusters, or spatial autocorrelative methods for understanding stochastic patterns.

\subsection{Galaxy clusters}

The strongly clustered spatial distribution of galaxies was recognized from galaxy counts of wide-field photographic plates during the mid-20$^{th}$ century. Many prominent astronomers conducted studies on this problem (including Edwin Hubble, Harlow Shapley, Fritz Zwicky, Gerard de Vaucouleurs, and P.J.E.\ Peebles).  Several statistical approaches were taken \citep[see the review by][]{1971PASP...83..113D}. One early result was that the frequency distribution of galaxy counts in quadrats followed a lognormal distribution rather than the Poisson distribution expected from spatial randomness. An ``index of clumpiness,'' the ratio of observed to expected variance in number counts, was investigated. The spatial autocorrelation function was found to have signal out to several degrees in the sky. \cite{1954AJ.....59..285S} produced contour maps of equal surface density based on a uniform kernel, remarking that ``So many aggregations stand out prominently that one is tempted to speculate that clustering may be a predominant characteristic of nebular [galactic] distribution.'' Shane teamed with Berkeley statisticians Jerzy Neyman and Elizabeth Scott to develop statistical models (such as a double Poisson model) of the distribution \citep{1953ApJ...117...92N}.

Some analyses can be viewed as mixture models for the galaxy distribution in the nearby universe, although they are not usually described in this way. \cite{1958ApJS....3..211A} conducted a heroic visual survey of the Palomar Observatory Sky Survey plates to identify several thousand individual galaxy clusters using a decision tree applied to visual galaxy counts. \cite{1976ApJS...32..409T} constructed a catalog of galaxy groups based on a single-linkage agglomerative clustering algorithm. (Their procedure became very popular in the astronomical community under the label ``friends-of-friends'' or percolation algorithm without awareness of its widespread use in other fields.) \cite{1987ApJ...321..280T} identified clouds, associations and groups of nearby galaxies in three dimensions from a dendrogram procedure with linkages based on the gravitational forces between galaxies.

Statistical approaches to galaxy clustering as a stationary stochastic process was initiated by P.J.E.\ Peebles in the 1960s based on the two-point (pair) correlation function and the Fourier power spectrum. These were particularly important as they were linked to the astrophysical theory of structure formation in an expanding universe \citep[see the review by][]{1979RvMP...51...21F}. For example, \cite{1986isos.book..212B} examined the correlation functions of peaks in 3-dimensional Gaussian random fields arising from gravitational attraction in an initial spectrum of weak density fluctuations arising from the radiation-dominated era after the Big Bang. Peebles' approaches are still in wide use today; for instance, the faint signal expected from Baryon Acoustic Oscillations, an important test for standard cosmological theory, was recently discovered using the two-point galaxy correlation function from SDSS data \citep{2005NewAR..49..360E}.

However, all of these early studies treated galaxy clustering as an isotropic process. But this assumption was radically invalidated as larger telescopes devoted observing time to galaxy redshift surveys. Redshifts represent an approximate measure of galaxy distance in the expanding universe and, when combined with location in the sky, give a 3-dimensional view of the galaxy distribution. When about 1000 redshifts were obtained, the distribution was found to resemble ``a slice through the suds in the kitchen sink.'' The language of galaxy clustering changed: ``clusters'' were now viewed as the intersections of ``filaments,'' ``sheets'' and ``Great Walls'' of galaxies that surround ``voids.'' The volume by \cite{2002sgd..book.....M} lays the foundation between 3-dimensional galaxy statistics and cosmological theory.

\begin{figure}[t!]
\begin{center}
\includegraphics[width=0.6\textwidth]{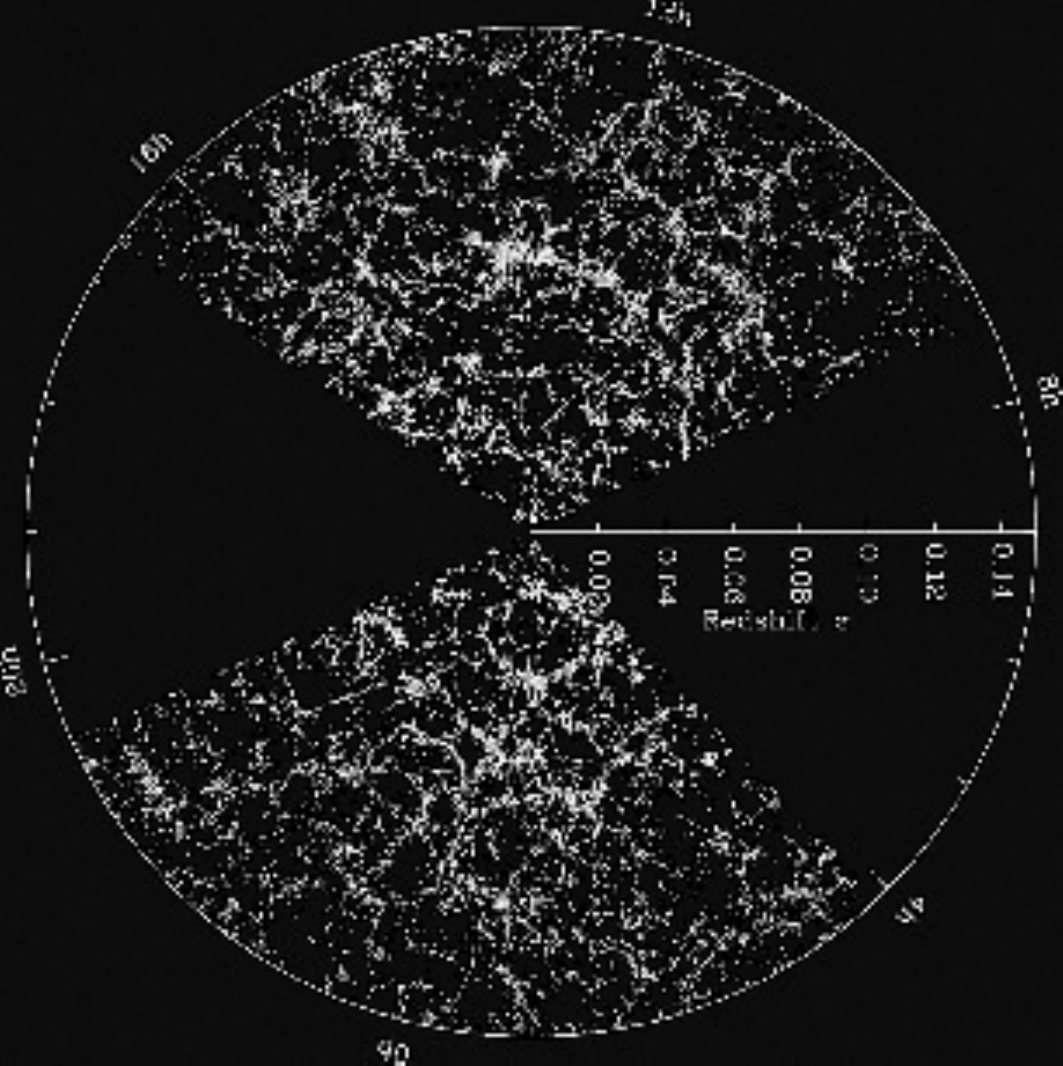}
\caption{Projection into 2-dimensions of a portion of the 3-dimensional Sloan Digital Sky Survey galaxy redshift survey showing the difficulties of mixture modeling of the ``cosmic web'' of galaxies in space.  https://www.sdss3.org/science/gallery\_sdss\_pie2.php\label{SDSS_galaxies.fig}}
\end{center}
\end{figure}

Increasing resources were devoted to constructing the 3-dimensional map of galaxies, most notably with the acquisition of more than 2 million galaxy redshifts by the SDSS \citep{2015ApJS..219...12A}. Figure \ref{SDSS_galaxies.fig} shows a two-dimensional projection of a small portion of this dataset. The links to cosmological theory are strong. Not only does the SDSS Fourier power spectrum agree well with the standard $\Lambda$CDM cosmological model \citep{2004ApJ...606..702T}, but massive simulations of structure formation in a Dark Matter dominated expanding universe accurately reproduce the soap bubble or ``cosmic web'' appearance of the galaxy distribution \citep{2005MNRAS.364.1105S}. Examination of the structure of galaxy clusters using mixture models has shown that most have clumpy, complex structures \citep{2012A&A...540A.123E}.

In light of these developments, it is not clear that mixture models can play a significant role in the characterization or understanding of the galaxy distribution. It is not clear either that the concept of distinct galaxy ``clusters'' is meaningful. Simple one-dimensional treatments of small datasets, as examined by a number of statisticians \citep{Roeder90,Escobar95,Carlin95,Phillips96,McLachlan97,Roeder97,Richardson97,Aitkin01,Aitkin11}, are no longer appropriate. The multi-scale, high-amplitude, anisotropic, web-like structure of the 3-dimensional galaxy clustering is difficult to treat using standard methods of mixtures, multivariate analysis, or spatial point processes. A number of heuristic algorithms for finding filaments or voids are in use, but with little foundation in statistical theory. There is thus a real need for development of statistical tools -- such as two-sample tests for comparing observations with simulations of different cosmological models -- that are well-suited to the complexities of galaxy clustering.

\subsection{Young star clusters}

Star formation is another topic in astronomy where the spatial clustering of objects, in this case young stars, can have important implications. Star formation is an ongoing process in many galaxies, including within our own Milky Way Galaxy, where the current rate is approximately 1 star per year \citep{2010ApJ...710L..11R}. Stars form in molecular clouds -- the coldest, densest phase of interstellar gas, which are mostly composed of H$_2$ -- when these clouds becomes unstable to gravitational collapse and contracts to form stars.  However, gravitational collapse must compete with phenomena that resist collapse, such as cloud turbulence and thermal and magnetic pressures. Thus, star formation is restricted to the densest cloud cores, within which stars form in groups that often merge into temporary rich clusters \citep{2003MNRAS.339..577B}. On galactic scales, new-born stars are concentrated in large complexes known as star-forming regions, which often lie within the spiral arms in many galaxies. These complexes last for several million years, after which most of the stars will have dispersed into the galaxy. However, some of the stars may remain in gravitationally bound groups known as open clusters.

\begin{figure}[t!]
\centering
\includegraphics[width=0.45\textwidth]{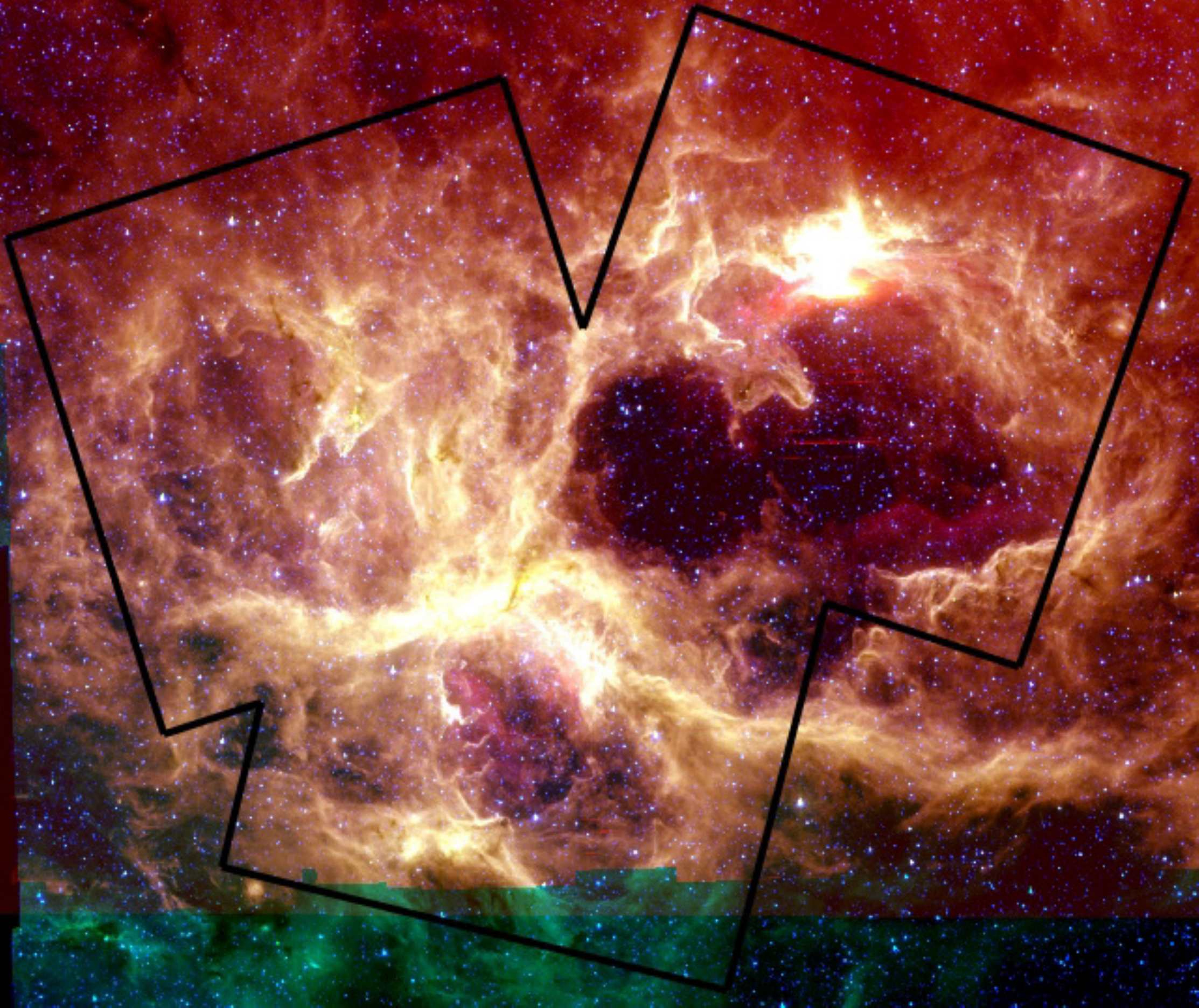}
\includegraphics[width=0.45\textwidth]{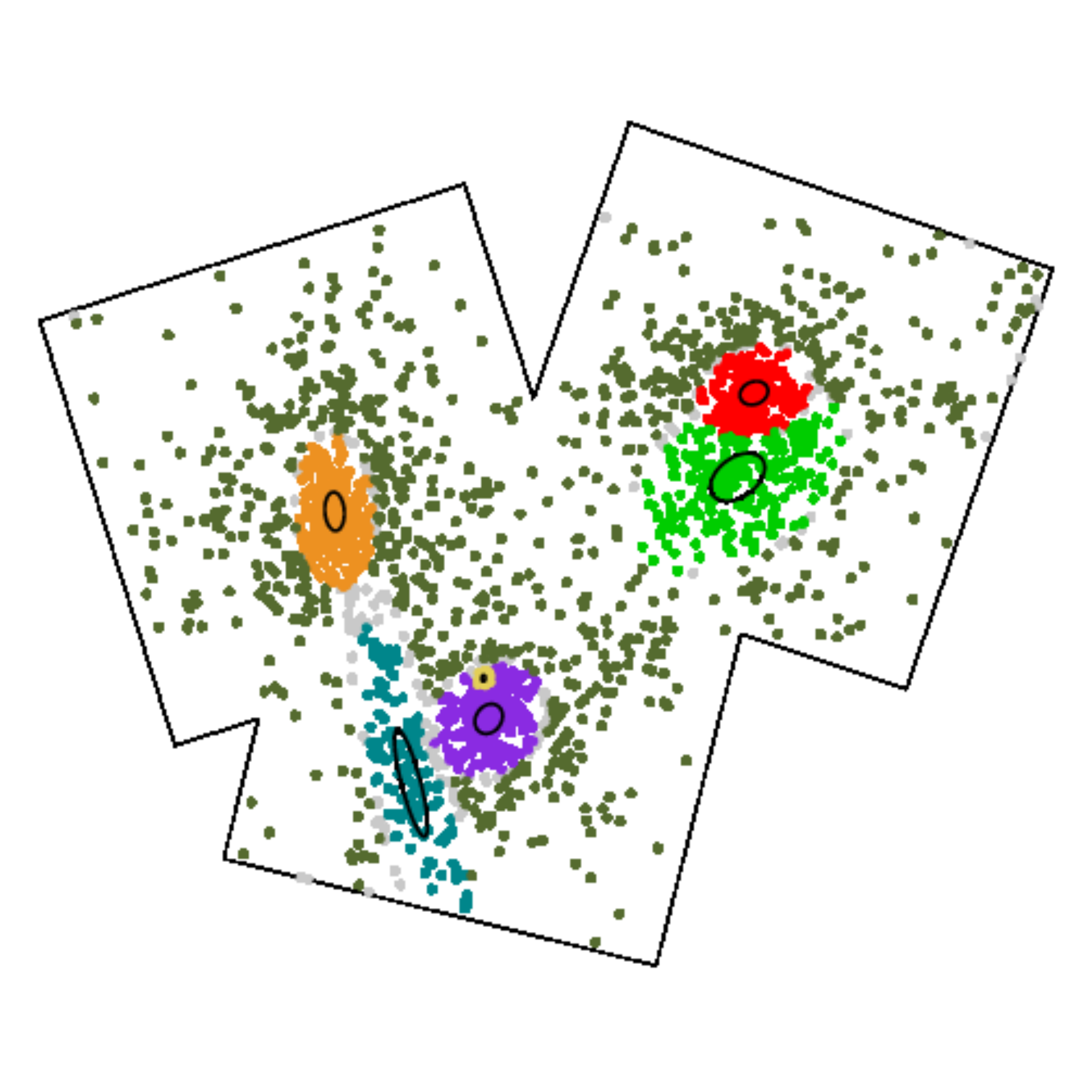}
\caption{Left: The mid-infrared view of NGC~6357 seen by the {\it Spitzer Space Telescope}. The molecular clouds, forming several bubbles, are seen prominently in these images but the star clusters are not immediately evident. Right: The cluster members identified from the X-ray/infrared MYStIX study, which are color-coded by group from the mixture model. Light grey stars have ambiguous cluster memberships, and dark green stars are members of a distributed  population in the model. The  core-regions of the various mixture components are shown as black ellipses.  \citep{2013ApJS..209...26F,2013ApJS..209...32B,2013ApJS..209...29K,2014ApJ...787..107K}
\label{mir.fig}}
\end{figure}

In star-forming regions, the locations of stars reflect the structure of the natal molecular cloud,
with the exact relation between cloud and star properties  a matter of active research \citep{2013ApJ...778..133L}. Gravitational collapse of the clouds causes the gas in star-forming regions to collapse and fragment into multiple clumps and filaments, so new born stars will typically be distributed in several subclusters within a star-forming region.  An example region is shown in Figure~\ref{mir.fig} -- the image from NASA's {\it Spitzer Space Telescope} shows both gas clouds and stars, while the spatial distribution of the stars, selected using data from NASA's {\it Chandra X-ray Observatory}, reveals multiple subclusters.

Individual stars can be identified and their spatial distributions analyzed in a number of star-forming regions in the Galaxy within a distance of several kiloparsecs -- a section of the Galaxy that includes part of our own spiral arm and neighboring spiral arms.    In this chapter, we discuss star clusters in 18 different star-forming regions included in the MYStIX study \citep{2013ApJS..209...26F} and related studies \citep{2014ApJS..213....1T,2017AJ....154...87K}. The mixture model analysis for star clusters was performed by \citet{2014ApJ...787..107K}, and similar methods were used by \citet{2017AJ....154..214K} and Getman et al. (2018, submitted). For the mixture model analysis, \citet{2014ApJ...787..107K} only used the two variables, right ascension (RA) and declination (Dec), which describe the stars' angular coordinates on the sky. The data are also are limited by the irregular fields of view observed by the {\it Chandra} telescope.  Information about the third radial dimension of stellar positions in the clusters is also not available.  In this example, we will refer to these two spatial coordinates as $x$ and $y$.

\subsubsection{Star-cluster models}

Older star clusters that have reached a quasi-equilibrium dynamical state are relatively well-understood, but young star clusters, which are still affected by the initial conditions of star formation, are not. Several families of spherically symmetric model have been used to fit the density profiles of star clusters. These include the isothermal sphere, the King profile, and the Plummer sphere,   which are all approximations to quasi-equilibrium distributions of stars in gravitationally bound groups \citep{2008gady.book.....B}.

For cluster analysis, \citet{2014ApJ...787..107K} used the isothermal sphere, which has been shown to provide a good empirical description of the distribution of stars in some young stellar clusters \citep[e.g.,][]{1998ApJ...492..540H,2008ApJ...675..464W,2014ApJ...787..107K,2017AJ....154..214K}. The model is unphysical at large distances from the cluster center because the number of stars diverges when integrated over all space. Nevertheless, this model provides an adequate fit to clusters within the observed fields of view.

The isothermal sphere has a characteristic ``core'' radius, $r_c$, which defines the size of the cluster. The distribution of stars projected in 2 dimensions on the sky (the surface-density distribution) can be approximated out to several core radii by an analytic expression known as the Hubble model,
$$
f_h(R)=\frac{A}{1+(R/r_c)^2},
$$
where $A$ is a constant and $R$ is the distance from the center of the cluster. However, many young stellar clusters are not spherically symmetric, but show significant ellipticity \citep{1998ApJ...492..540H,2017AJ....154..214K}. Generalization of this model to allow for elliptical contours of equal density requires the introduction of two new model parameters -- ellipticity $\epsilon$ and the ellipse orientation $\varphi$ on the sky. The resulting surface density for the \lq\lq isothermal ellipsoid\rq\rq\ at the coordinates $r=(x,y)$ is described by the equation
\begin{multline}
f_\mathrm{ie}({r};A,x_0,y_0,r_c,\varphi,\epsilon) = \\
A \left [1+
\left| \left[ \begin{array}{cc}
(1-\epsilon)^{-1/2}\cos \phi & (\epsilon-1)^{1/2}\sin \phi \\
(1-\epsilon)^{-1/2}\sin \phi& (1-\epsilon)^{1/2}\cos \phi \\
\end{array} \right]
\left[
\begin{array}{c}
\Delta x\\ \Delta y \\
\end{array}
\right]
\right|^2\middle/r_c^2
\right]^{-1},
\end{multline}
where ${r_0}=(x_0, y_0)$ is the center of the ellipsoid, and $\Delta x = x-x_0$ and $\Delta y=y-y_0$.  Thus, this cluster-component model has six parameters: $x_0$, $y_0$, $r_c$, $\epsilon$, and $\varphi$ and a normalization parameter $A$. \citet{2014ApJ...787..107K} call this model the ``isothermal ellipsoid.'' However, this model is meant merely as an empirical description of the projected spatial distribution of stars in a cluster, since information about a cluster's dynamical states is lacking.   Furthermore, this model is only applicable to the field of view observed by the telescope, which we denote the window $W$.

\begin{figure}[t!]
\centering
\includegraphics[width=0.8\textwidth]{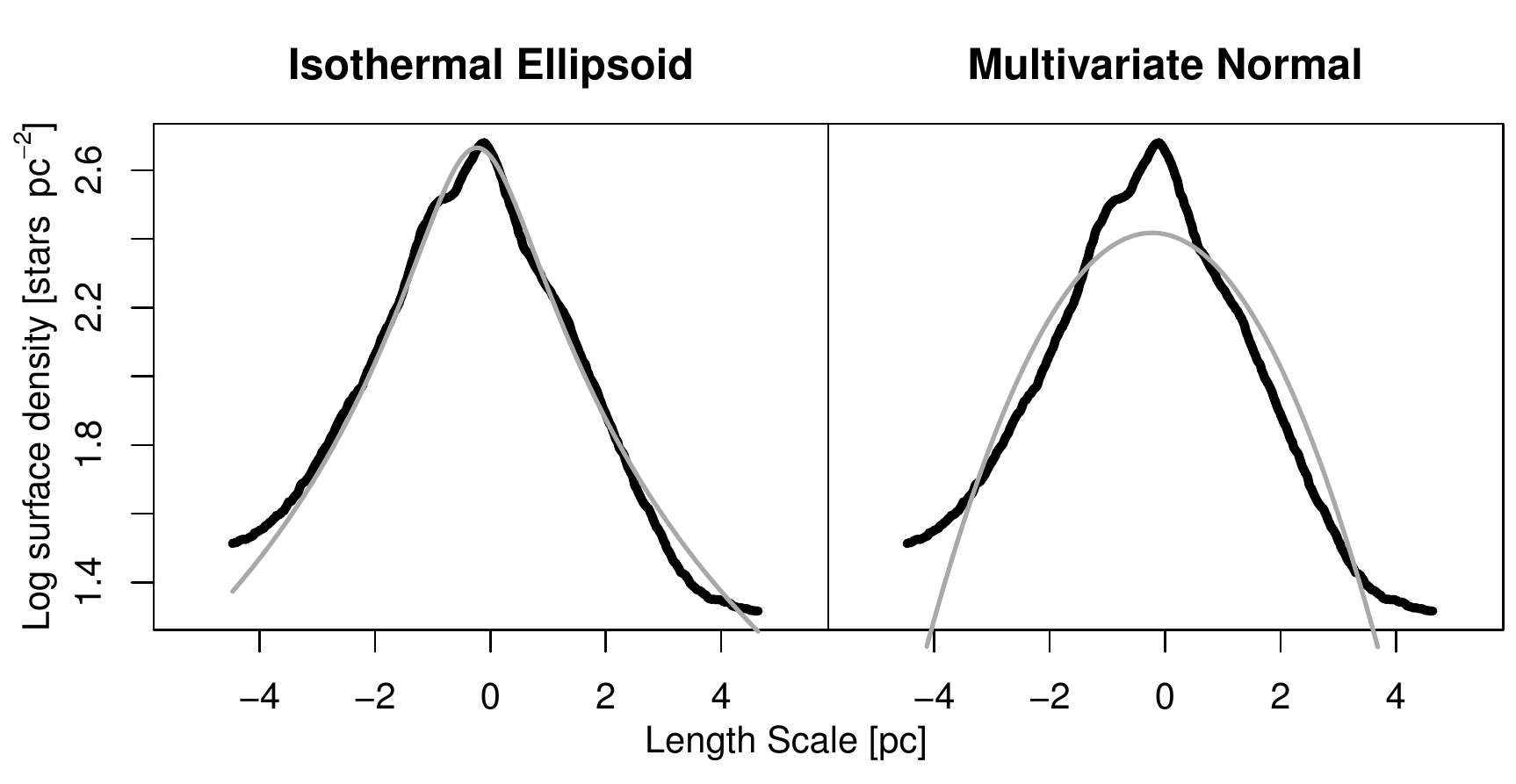}
\caption{The black lines in each plot show density of stars in the region NGC~6231 estimated through adaptive smoothing. The gray lines show the models that have been fit. The left panel shows the ``isothermal ellipsoid'' model and   the right panel shows the normal distribution model. (The $y$-axes of these plots are shown with a logarithmic scale.) Clearly, the isothermal ellipsoid model provides a better description of the data. \citep{2017AJ....154..214K}
\label{cuts.fig}}
\end{figure}

 The \lq\lq isothermal ellipsoid\rq\rq\ model provides a closer match to observed young stellar clusters than other, better understood, distributions like the multivariate normal distributions or Student's $t$-distribution.
Figure~\ref{cuts.fig} shows the cluster NGC~6231 fit with two models, the isothermal ellipsoid model on the left and a multivariate normal distribution on the right. These models show a slice through the two-dimensional surface density maps, with the non-parametrically smoothed data shown in black and the models shown in gray.  Surface densities [ordinate] are shown with logarithmic values.  While the isothermal ellipsoid provides a good match to the data with only minor deviations at large distances from the center, the multivariate normal model misses both the cluster center and the wings of the distribution. When using normal functions to fit stellar surface density distributions, the modeling tries to compensate for this mismatch by using several, approximately concentric, normal mixture models to fit a single star cluster.

In addition to the \lq\lq isothermal ellipsoid\rq\rq\  components, \citet{2014ApJ...787..107K} used an additional component, $f_{\mathrm{U}} ({r})$, to model stars distributed uniformly in the field of view.   These can either be young stars that are not part of clusters (e.g., a ``distributed population'') or contaminants in the sample which are expected to exhibit complete spatial randomness. This approach is also used by the well-known normal mixture model procedure {\sf mclust} to deal with noise and outliers \citep{mclust12}.

The mixture model for  the spatial distribution of the stellar population will be the sum of the isothermal-ellipsoid models for $G$ clusters plus the unclustered component,   each of which is weighted by mixing coefficients, $\eta_g$. This model is given by the equation,
$$
p({r}| \theta) = \sum_{g=1}^{G+1}\eta_g f_{g}({r}|\theta_g) = \sum_{g=1}^{G}\eta_g f_\mathrm{ie}({r}|x_{0,g},y_{0,g},r_{c,g},\varphi_g,\epsilon_g)+\eta_{G+1} f_{\mathrm{U}} (r),
$$
where $\theta = \{\eta_1,x_{0,1},y_{0,1},r_{c,1},\varphi_1,\epsilon_1,\ldots, \eta_G,x_{0,G},y_{0,G},r_{c,G},\varphi_G,\epsilon_G,\eta_{G+1}\}$ denotes the model parameters.
The model thus has six parameters for each ellipsoidal component, $\theta=(x_0, y_0, r_c, \epsilon, \varphi)$ and one for the uniform component mixing parameter   but one fewer degree of freedom because the model must be normalized, yielding $6G$ dimensions for the full model.

\subsubsection{Model fitting and validation}

The log-likelihood for a point pattern within a finite window ($W$) is given by the equation
\begin{equation}\label{likelihood}
\ell_o  (\theta) = \log p(r_1,...,r_n| \theta)=\sum_{i=1}^{n} \log  p({r}_i| \theta) ,
\end{equation}
under the assumption that the pattern of points, $\{r_1,\ldots ,r_n\}$, is generated by an inhomogeneous Poisson point process containing $n$ points. The mixture model $p({r}| \theta)$ must be normalized   in the window $W$, which is done by numerical integration due to the irregular shape of the window.  The CRAN R package {\sf spatstat} also makes use of irregular windows for analysis of spatial point processes \citep{baddeley2015spatial}.

\citet{2014ApJ...787..107K} carried out the maximum likelihood estimation (MLE) by directly searching for the maximum of $\ell_o  (\theta)$, rather than using the typical Expectation--Maximization (EM) approach. This method was used because there is no ML formula for the parameters of the isothermal-ellipsoid model for points within an irregular window $W$.  Direct searching can be computationally challenging   due to the high dimensionality of the parameter space. While standard optimizers like the EM algorithm treat all of the variables equally, in this case the scientific motivation requires that the clusters be present in two variables $(x,y)$ while the other variables ($r_c$, $\varphi$, $\epsilon$) are secondary. \citet{2014ApJ...787..107K} started the MLE computation with a superset of possible clusters obtained from bumps in an adaptively smoothed surface-density map of the point process.  A Nelder-Mead optimization algorithm, implemented in $R$'s function {\it optim}, was used to find the global maximum likelihood. In $R$, even for complicated distributions, with $n\approx1000$~stars and $G \approx 10$ cluster components, the Nelder-Mead algorithm produces a reasonable solution in less than 15 CPU-minutes.

\citet{2014ApJ...787..107K} based model selection on minimizing the Akaike Information Criterion (AIC),
$$
\mathrm{AIC} (G)=-2 \ell_o  (\hat{\theta}_G) +2(6G),
$$
see Chapter~7, Section~7.2.2 for a review of   information criteria for model selection.
Although there has been much debate over which penalized likelihood to use for model selection \citep[e.g.,][]{Lahiri01,Konishi08,Burnham02,KassRaftery95}, for this problem the AIC has several advantages.  A typical star-forming region may have a large dynamic range in the numbers of stars in young stellar clusters. For example, a subcluster of $\sim$20 stars may reside next to a rich cluster with $\sim$500 stars.
In addition, clusters may be superimposed on each other, either due to the projection of multiple discrete clusters along the same line of sight or astrophysical cases of core--halo structure.  The AIC has greater sensitivity at probing these effects than, say, the Bayesian Information Criterion (BIC).

\begin{figure}[t!]
\centering
\includegraphics[width=0.90\textwidth]{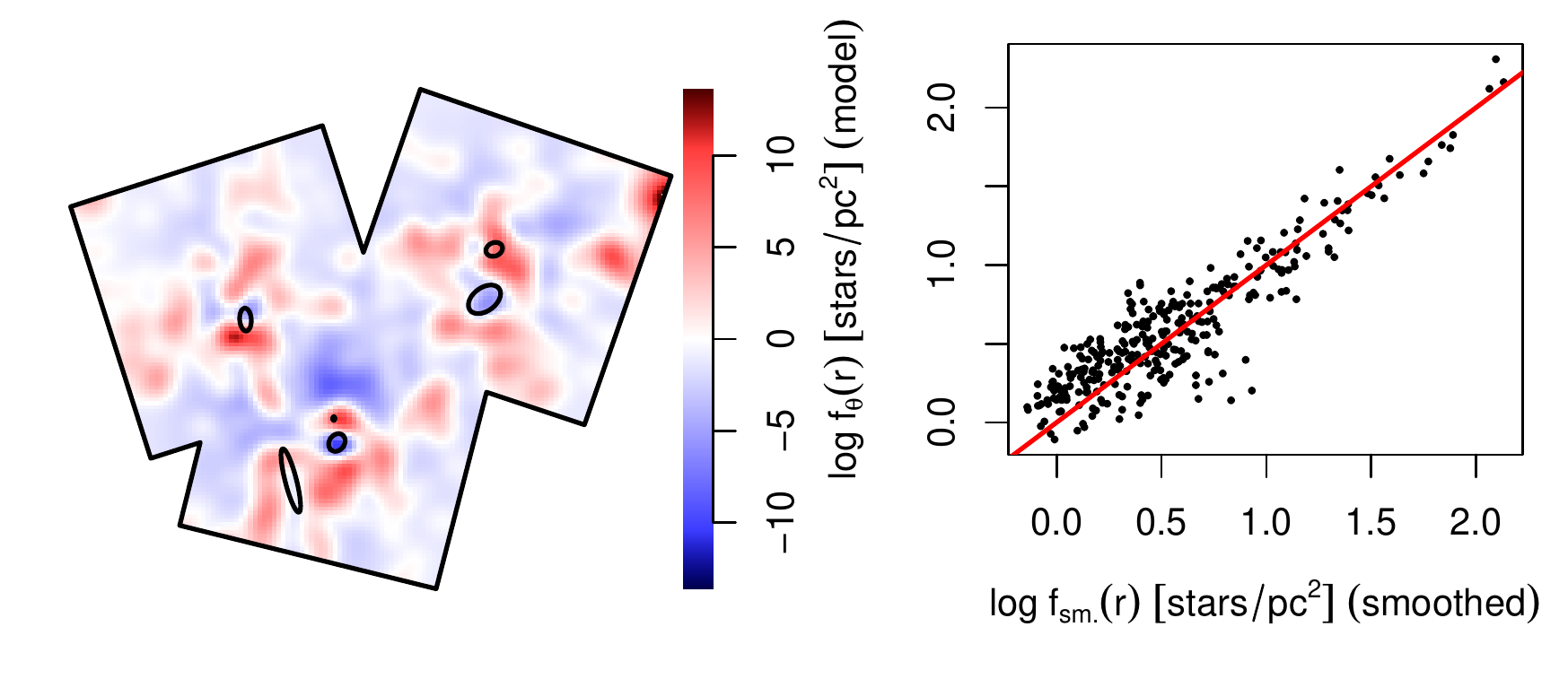}
\caption{Left: Residual surface density for NGC~6357. Negative residuals are shown in blue and positive residuals are shown in red. The peak residuals are roughly $\sim$10\% of the peak surface density in a smoothed map of the observed star distribution. Right: Density obtained from the mixture model is plotted against  density from the adaptively smoothed surface density maps. \citep{2014ApJ...787..107K} \label{resid.fig}}
\end{figure}

To validate the accuracy of the mixture model MLE, \citet{2014ApJ...787..107K} examined kernel smoothed residuals of the mixture model.  The construction of these residual maps is described by \cite{Baddeley05,Baddeley08} and implemented using the {\it diagmose.ppm} tool in the  software R package
{\sf spatstat} for statistical analysis of spatial point processes. Residual maps can indicate both the amplitude of residuals and give insight into physical deviations from the model assumptions.  Better fits will have lower amplitudes and lack coherent structures in residual maps. The identification of possible missing clusters using the residual maps is similar to the use of ``final prediction error'' to fit a model recommended by \cite{Rao01}. Figure~\ref{resid.fig} shows the kernel smoothed residual map for the star-forming region NGC~6357, while the previously mentioned Figure~\ref{cuts.fig} shows a comparison between the model prediction and the smoothed data in the young stellar cluster NGC~6321.

\subsubsection{Results from the mixture model approach}

Several nonparametric methods have been used in the astronomical literature for identifying clusters of stars in star-forming regions based on the minimal spanning tree, Voronoi tessellations, kernel density estimation, and nearest neighbor distributions \citep{2011AN....332..172S}.  But the parametric mixture model approach offers a decisive advantage: an astrophysical model from which astrophysical inferences may be devised. Three important quantities obtained from the isothermal ellipsoidal fit to each subcluster are the core radius of the cluster (in parsecs), the central star density (in stars per cubic parsec), and the total number of stars in the cluster. None of these parameters may be obtained from nonparametric methods: the core radius is not clearly defined without a model, inferring 3-dimensional properties from 2-dimensional data requires  a model, and overlapping clusters impedes counting the number of stars in a cluster.

\begin{figure}[t!]
\centering
\includegraphics[width=0.45\textwidth]{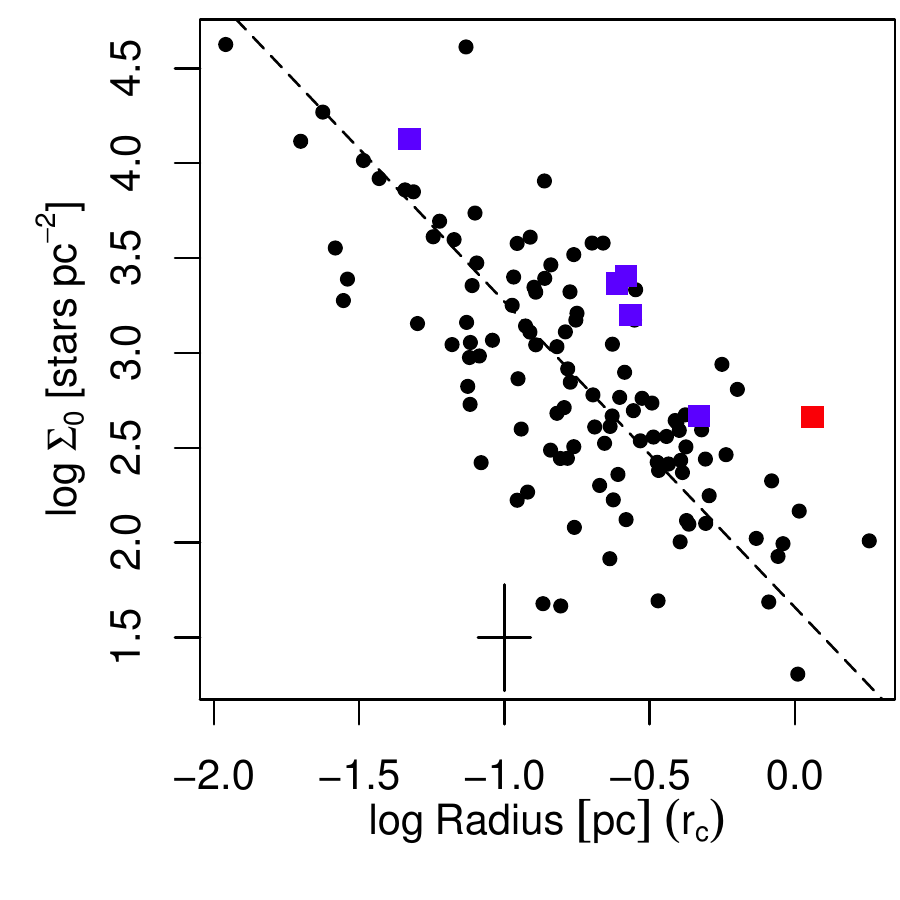}
\caption{Scatterplot showing properties of clusters identified by the mixture model analysis in 18 star-forming regions. The ordinate is the density of stars at the center of a cluster, $\Sigma_0$, and the abscissa is the core radius of the cluster, $r_c$, both of which are derived from the mixture model. Blue points are clusters from NGC~6357 and the red point is from NGC~6231. The black cross indicates typical 1$\sigma$ uncertainties on the model's parameters.
\citep{2017AJ....154..214K} \label{compare.fig}}
\end{figure}

Figure~\ref{compare.fig} shows, for the full set of 18 star-forming regions, the relations between two of these variables: $\Sigma_0$ -- the density of stars at the center of a cluster; and $r_c$ -- the cluster's core radius. This plot is of astrophysical interest because it can be interpreted with astrophysical models about how star clusters form and evolve. For example, decreasing density with increasing radius can be interpreted as an effect of cluster expansion \citep{2009A&A...498L..37P}. Subsequent investigation of the ages of stars in these clusters support this interpretation \citep{2014ApJ...787..108G,2015ApJ...812..131K,2017AJ....154..214K}.

\section{Classification of astronomical objects}

Classification is very important to astronomers: a full-text search of the astronomical literature published in 2015 shows that 30\% of all papers are concerned with classification. Using modern observatories, it is common for astronomers to gather data from a large number of objects of different types. This can arise from collections of images containing thousands of stars or galaxies in a single field of view, or from fiber-fed spectrographs capable of obtaining dozens of spectra simultaneously.  Imaging and spectroscopic observations provide information   that can be used  to group objects with similar properties including photometry  (the brightness of objects at various wavelength bands), spectroscopy (showing emission and absorption lines from atoms and molecules that are signatures of chemical, thermal, and electromagnetic properties), time-variability data, morphology from structure in the images, and motions on the sky.

Different classes are sometimes clearly distinguished. For example, gamma-ray bursts have been subdivided into two classes, long- and short-gamma ray bursts, based on distinct peaks in the distribution of prompt burst durations, with a relatively clear division at 2 seconds \citep{1993ApJ...413L.101K}. However, in other cases, there can be significant overlap in the classes' properties and it may not be possible, with limited data, to reliably classify an object.

 One of the first problems in astronomy that used the concept of mixtures was groups of co-moving stars in the Galaxy. These groups, now called stellar associations or moving groups, are important because they can represent groups of stars that all formed in one star-formation episode. In an investigation of star streams in the Galaxy with different kinematic properties (now recognized to be based on spurious data), \citet{1906MNRAS..67...34E} wrote:
\begin{quote}
``We cannot, as a rule, pick out an individual star and decide (from its motion) to which drift [moving group] it belongs. However, we may roughly separate out stars typical of the two drifts and examine their characteristics."
\end{quote}
Mixture models are commonly used in astronomy for distinguishing a single class of objects from a mixture of multiple classes of objects. The most-cited papers on mixture models in astronomy by \cite{1994AJ....108.2348A} and \cite{2010ApJ...718.1266M} describe methods for hypothesis tests to distinguish a unimodal normal distribution from mixtures of multiple normal distributions.  Astronomers may interpret the empirical appearance of multimodality as evidence for physically distinct classes, and this is often confirmed through follow-up studies.

\subsection{Tests for multiple components}

A variety of tests for multimodality are used by astronomers, typically based on assumptions of multivariate normal distributions.   This often involves the classical likelihood ratio test, AIC, BIC, or full calculation of the Bayes factor.  The null hypothesis of a single class can be rejected in favor of an alternate hypothesis of a mixture of multiple classes  \citep[e.g.,][]{Jeffreys61,KassRaftery95}. However, there is no consistency on the choice of AIC, BIC, or other model selection approach within the astronomical literature.

 The use of normal mixture models to perform a hypothesis test of bimodality by \cite{1994AJ....108.2348A} has been influential in the astronomical literature. They test whether univariate data, $\{y_0, \ldots ,y_n\}$, are consistent with a normal distribution (the null hypothesis) or are better described by $G$ equal-width normal distributions (the alternate hypothesis).
For the $G$-component model, the pdf and the   complete-data  log-likelihood are given by the equations,
\begin{eqnarray*}
p(y|\mu,\sigma^2) &=& \sum_{i=1}^G \eta_g\phi(y|\mu_g,\sigma_g^2) ,\\
\ell_o (\mu,\sigma^2) &=& \sum_{g=1}^{G}\sum_{i=1}^{n}z_{gi}(\log \eta_g+\log \phi(y_i|\mu_g,\sigma_g^2)),
\end{eqnarray*}
where the $\mu_g$'s are the component means, $\sigma_g$ are the standard deviations, the $\eta_g$'s are the component mixing parameters, and the $z_{gi}$'s are indicator variables with value 1 when the $j$th point is assigned to group $g$ and 0 otherwise. The assignments of points to groups are not known {\it a priori}, so the EM algorithm is used, as presented in Chapter~2.
We recall here that, during the  E~step the $\hat{z}_{gi}$ are calculated using
\begin{equation*}
\hat{z}_{gi}=\eta_g \phi(y_i|\mu_g,\sigma_g^2)/p(y_i|\mu,\sigma^2).
\end{equation*}
Then, during the M~step maximum likelihood values of mixing parameters, $\eta_g$, the group means, $\mu_g$, and the common variance of the groups, $\sigma_g^2$, are obtained with the following equations:
\begin{eqnarray*}
  \label{m1.eqn}  \hat{\eta}_g &= &\frac{1}{n}\sum_{i=1}^{n}\hat{z}_{gi}, \\
  \label{m2.eqn}   \hat{\mu}_g&=&\frac{1}{\hat{\eta}_g n}\sum_{i=1}^{n}y_i\hat{z}_{gi}, \\
  \label{m3.eqn} \hat{\sigma}_g^2&=&\frac{1}{\hat{\eta}_g(n-1)}\sum_{i=1}^{n}(y_i - \mu_g)^2\hat{z}_{gi}.
\end{eqnarray*}
For homoscedastic cases where all components have a common variance, the standard deviation in each M~step can be obtained from a weighted mean of the  $\hat{\sigma}_g^2$ values,
\begin{equation*}
\hat{\sigma}^2=\sum_{g=1}^G\eta_g \hat{\sigma}_g^2.
\end{equation*}
\cite{1994AJ....108.2348A} used the log-likelihood ratio test statistic, $-2\log \lambda =-2(\ell^ 1-\ell^G)$, to test for statistical significance. For mixture models in general, $-2\log \lambda$ does not have its usual asymptotic $\chi^2$ distribution due to a breakdown in the regularity condition in the case of finite mixture models \citep{GhoshSen1985}, see also Chapter~7, Section~7.2.1.
 However, for mixture models where the variances of all components are equal, \cite{Wolfe71} found empirically that a statistic $-2C\ln \lambda$ approximately follows a $\chi^2$ distribution. With $d$-dimensional data modeled with $G_0$ (null hypothesis) and $G_1$ (alternate hypothesis) components, the correction factor is $C=(n-1-d-G_1/2)/n$ and the number of degrees of freedom of the $\chi^2$-distribution is
equal to $2d(G_1-G_0)$.  \cite{1994AJ....108.2348A} were unaware of this subtlety and report $p$-values based on the $\chi^2$ distribution with the approximation $C=1$.

A statistically significant identification of multiple components is not sufficient to identify multimodal distributions, because mixtures of closely spaced components may have a single mode, as discussed in Chapter~1. \cite{1994AJ....108.2348A} used the statistic
\begin{equation}
\Delta=\frac{|\mu_1 - \mu_2|}{\sigma},
\end{equation}
to measure the separation between components. In the case of equal mixing parameters, a two component distribution will be multimodal when $\Delta>2$. Hartigan's dip test may be used to test for the existence of bimodality of a dataset \citep{Hartigan85}. However, \cite{2010ApJ...718.1266M} argue that, for bimodal distributions with equal mixing parameters and the same value of $\Delta$, the log-likelihood ratio test is more sensitive to multiple normal distribution components than the dip test is to bimodality. Nevertheless, in astronomy, the identification of multiple components is often more scientifically interesting than the identification of bimodality in a particular set of variables \citep{2015MNRAS.446.2144T}.

 \cite{2010ApJ...718.1266M} presented   a new, more general, code for establishing the presence of heteroscedastic mixture components. Three statistics are investigated: the log-likelihood ratio test statistic, the kurtosis of the distribution (a negative kurtosis is a necessary condition of a bimodal distribution produced by a two-component normal mixture model), and the separation between the two components, now defined as
\begin{equation*}
\Delta=\frac{|\mu_1 - \mu_2|}{\sqrt{(\sigma_1^2 + \sigma_2^2)/2}}.
\end{equation*}
Once the model is fit with the EM algorithm, non-parametric bootstrap resampling is performed to estimate uncertainties on the model parameters and $\Delta$. Finally, the parametric bootstrap is run to estimate the $p$-value for a unimodal distribution.

The statistical tests provided by these codes have been used in many astronomical studies.  Most commonly, this test is used to study the distribution in the color of astronomical sources.  In astronomy, color refers to the ratio of the amount of light observed in one band (e.g., the $V$ band) to the amount of light in another band (e.g., the $B$ band). Brightness is usually measured in magnitudes, a logarithmic unit where larger values indicate dimmer sources, so a color would be written as $B-V$, with a larger value indicating relatively more light in the $V$ band than in the $B$ band. Hundreds of papers have referenced these codes in investigations of colors of stars, globular clusters, and galaxies \cite[e.g,][]{2007AJ....133.1447W,2001AJ....121.2974L,2001AJ....121.2950K}.

We now present three individual studies that illustrate a variety of common characteristics of astronomical classification problems.

\subsection{Two or three classes of GRBs?}

The two major classes of gamma-ray bursts (GRB), short and long GRBs, were identified by a distinct bimodality in the distribution of burst durations \citep{1993ApJ...413L.101K}. Today, there is ancillary evidence that the classes are physically distinct: long-duration gamma-ray bursts are produced by collapsars, the implosion of a massive star at the end of its life, while short-duration gamma-ray bursts originate from the merger of binary neutron stars \citep{2007PhR...442..166N,2017ApJ...848L..12A}.  The early analyses revealing these two classes were based on univariate or bivariate distributions, but observations of GRBs provide a larger variety of properties that can be used for classification.

 \cite{1998ApJ...508..314M} examined the clustering of GRB properties using a larger set of variables than were typically included in previous studies to provide a fuller picture of the classes of GRBs. Their sample consisted of 797 GRB events from the BATSE instrument on NASA's {\it Compton Gamma-Ray Observatory}, each of which is described by 15 variables. This analysis suggested the presence of a third group of GRBs with intermediate properties.

The variables in the study included several measures of burst duration, fluence (the total amount of light observed from the burst), and spectral hardness (the average energy of the observed photons). Analysis of these variables showed that several of them are redundant: some mainly add noise to the clustering process, while others were not astrophysically meaningful.  A reduced set of five variables was obtained.

Cluster analysis was performed using two methods: hierarchical average-linkage clustering and normal mixture models. The cluster analysis used unit-free variables obtained through logarithmic transformation of the variables. This practice is more natural for use in astronomy than standardization by the sample standard deviation, since measurements often vary by several powers of ten, and in many cases measurements can be approximated by log-normal distributions. The hierarchal clustering was performed using a Euclidean distance metric (which is unit-dependent), and the number of clusters was selected based on the squared correlation coefficient (the fraction of the total variance accounted for by a partition into $G$ clusters) and the squared semi-partial correlation coefficient (the difference in the variance between the resulting cluster and the immediate parent clusters normalized by the total sample variance). Each analysis approach suggested that three classes of bursts are present: Class~I (long, bright, soft bursts), Class~II (short, faint, hard bursts), and Class~III (intermediate, intermediate, soft bursts). The first two categories reproduce the long GRB and short GRB classes, but the third class was new.

The mixture model analysis was performed using the  {\sf mclust} software \citep{Fraley98} for normal mixture models where the number of clusters is evaluated using the BIC. The set of variables were further reduced to three for this analysis. The best value of BIC was found for three clusters, with the difference between 2 and 3 clusters being $\Delta\mathrm{BIC}=68$, strongly supporting the results from hierarchical clustering that more than two classes of GRB exist.

It is still not certain whether this third class of GRBs is an astrophysically distinct class.  NASA's more recent {\it Swift} Gamma Ray Burst and $Fermi$ missions have discovered most known GRBs. Evidence for this third class of GRB has been weaker or absent in these later samples.  For example, a multivariate analysis of GRBs detected by {\it Fermi} finds that a two-component mixture model is highly favored \citep{2016ApJS..223...28N}. It is possible that the presence of a third component in the original dataset may have been an effect of sample selection caused by uninteresting properties of the BATSE instrument, rather than a distinct astrophysical class.

\subsection{Removal of contaminants}

It is often desirable to obtain a large samples of astronomical objects of a particular type. However, source lists obtained from observations may include contaminants, which are objects of a different type that masquerade as objects of the desired class. It is often difficult to completely eliminate contaminants from a large study without extensive followup observations, but in many cases some level of contamination is acceptable if the contaminant rate is kept sufficiently low.

\cite{2009ApJS..180...54J} performed a study of globular star clusters within nearby galaxies observed by NASA's {\it Hubble Space Telescope} (HST).
 Sources of light were identified within the images taken by HST's ACS camera, which include globular clusters, as well as contaminant foreground stars and background galaxies. (The host galaxy, which contains the globular clusters was ignored in the analysis.) For each of the sources, photometric $g$ and $z$ magnitudes and a characteristic radius, $r_h$, were measured.
Foreground stars may be easily distinguished and removed from the catalogs because their radii in the image are nearly zero. Globular clusters typically have smaller radii and brighter $z$ magnitudes than background galaxies, but these two populations overlap. These observations were made for 100 host galaxies, each observation having its own population of globular clusters and background galaxy contaminants.

To distinguish between globular clusters and background galaxies, \cite{2009ApJS..180...54J} used a mixture model strategy. The distribution of globular cluster properties $(r_h, z)$ was assumed to be universal for all cases, and was taken from prior knowledge of the well-studied globular cluster properties. The only free parameters is the mean radius, $\mu_{r_h}$, of the clusters.  The distribution of background galaxy properties $(r_h, z)$ was estimated for each observation, using a separate ``control field'' near to the original observation on the sky. All the sources in these control fields (once stars are removed) were assumed to be background galaxies, and the distribution of these sources was estimated using kernel density estimation. This density distribution has no free parameter to be fit in the mixture model analysis.

The mixture models were fit using the EM algorithm to find the mixing parameters and $\mu_{r_h}$ for each of the 100 observations. This method has two main advantages over a more typical method used by astronomers of using a fixed boundary between globular clusters and background galaxies in $(r_h, z)$-space. The mixture method accounts for variation in size of globular clusters from one host galaxy to another. In addition, the soft classifications provided by the mixture models allow samples of probable globular clusters to be obtained with different levels of contamination, depending on the needs of different science questions.

\subsection{Red and blue galaxies}

Galaxies generally fall into two groups: one class known as early-type galaxies (which are smaller, older, redder, and less likely to have star formation) and the other class known as late-type galaxies (which are larger, younger, bluer, and more likely to have star formation). The presence of two classes can be seen in distributions in galaxy colors, for example the Sloan $g-i$ color index, with a group of ``blue'' galaxies and a group of ``red'' galaxies. The two populations have color distributions that overlap, which means that some ``red'' galaxies have bluer colors than some ``blue'' galaxies, and vice versa. A number of studies have dealt with this distribution using a line on the color-magnitude diagram to separate both classes, with objects falling on one side of the line being assigned to one class and objects on the other side assigned to the other class \cite[e.g.,][]{2003ApJS..149..289B,2004ApJ...600..681B,2010ApJ...721..193P}. However, when different proposed dividing lines are applied, the properties of the resulting samples, for example their galaxy mass distributions, will differ \citep[][Figures~3 and~4]{2015MNRAS.446.2144T}.

\cite{2015MNRAS.446.2144T} instead used a mixture model approach to this problem. Their galaxy data originated from the GAMA project \citep{2011MNRAS.413..971D}, from which they derived a subset containing more than 23,000 objects pruned for reliability and to avoid biases from selection effects. Two variables were included in the analysis, galaxy mass $M_\star$ and $g-i$ color (corrected for redshift).    A complex model with 40 parameters was used to describe the ``red'' and ``blue'' galaxies in ($\log M_\star$, $g-i$) space. This model used gamma functions (known in astronomy as the \citet{1976ApJ...203..297S} function) to describe the distribution of galaxy masses, and models describing the color--mass relations, the scatter around these relations, and outlying data points. This model has more parameters than are necessarily demanded by the data. However, the purpose of the model is to provide a sufficiently flexible model that will describe the data, not an in-depth study of the model parameters; and the authors stated that, based on their analysis, they were not ``grossly overfitting the data.'' Fitting was performed with a Markov chain Monte Carlo method with uniform or uninformative priors using the Python software package {\sf EMCEE} \citep{2013PASP..125..306F}.

The result of the analysis was a soft classification of galaxies into two populations, with no evidence for the existence of an intermediate ``green'' population of galaxies. For both the ``red'' galaxies and the ``blue'' galaxies, the mass functions were similar to single Schechter functions. Small deviations included an excess of ``red'' galaxies with low mass and a deficit of ``blue'' galaxies with high mass. The mixture model mass functions avoid some of the unexpected artifacts present in the mass functions produced by hard classification methods. The models of the two populations show that colors of ``blue'' galaxies do not depend strongly on mass, but that the colors of ``red'' galaxies do vary strongly with mass. The most massive ``red'' galaxies have red $g-i$ colors indicative of very little star formation. However, the mixture model fit suggests that these galaxies are only one end of a broader distribution, which includes lower-mass ``red'' galaxies that are not as different in color from the ``blue'' population \cite[][Figures~10 and~11]{2015MNRAS.446.2144T}.

\section{Advanced mixture model applications}

For more advanced statistical modeling of data, it is often convenient if a distribution can be described by a flexible parametric model, and mixtures of normal distributions are one such possibility.
Many probability density functions can be mimicked by normal mixture models \citep[see, for instance,][]{Mclachlan00}, so these models can often be used to estimate distributions of data even in cases in which there is no theoretical reason to suspect that the data should originate from multiple components. Here, we describe two applications in astronomy where advanced methods for dealing with problems such as heteroscedastic measurement errors or missing values become feasible when it is assumed that underlying distributions are described by normal mixture models. These examples demonstrate how a mixture model can be incorporated into a hierarchical statistical model, facilitating the computation of a likelihood for complicated scenarios that may arise in astronomy.

\subsection{Regression with heteroscedastic uncertainties}

Astronomers are often interested in the relationship between two properties of a cosmic population, but must infer results from samples that are limited by telescope sensitivity or subject to significant heteroscedastic measurement uncertainties.  Fortunately, measurement uncertainties can usually be directly measured from calibration tests conducted under identical conditions to the true observation, and can thus enter the dataset rather than be parametrized in the model.

A widely cited treatment of such problems in astronomy is a bivariate regression procedure involving semi-parametric density estimation using mixture models by \citet{2007ApJ...665.1489K}. This approach uses normal mixture models as part of a hierarchical model of the problem, a strategy developed by generalizing a model presented by \citet{carroll1999nonparametric} to allow for heteroscedastic measurement error. In this case, the mixture model is an internal part of an algorithm, rather than a fundamental property of the input or the output of the statistical analysis. Thus, the properties of the mixture model, such as number of components and component parameters, are not important so long as the model can provide an adequate approximation of the underlying distribution. By using a mixture model framework, \citet{2007ApJ...665.1489K} is able to construct a likelihood for a hierarchical model, which can then be used to perform maximum-likelihood estimation or Bayesian inference.

For a case of linear regression with measurement errors on both the independent and dependent variables,  \citet{2007ApJ...665.1489K} construct a hierarchical model. Note that in this chapter we have altered the notation used by \citet{2007ApJ...665.1489K} to be consistent with usage in the rest of the book. We denote the intrinsic value of the independent variable $\xi$ and the dependent variable $\iota$ with the relation
  \begin{equation*}
    \iota_i = \alpha + \beta \xi_i + \epsilon_i,
  \end{equation*}
where $(\alpha,\beta)$ are the regression coefficients and the error term $\epsilon_i$ is a random variable drawn from a normal distribution with variance $\sigma^2$. However, observational effects will yield measurement errors (which may be correlated) on both $\xi$ and $\iota$. We model the relation between observed values $(x,y)$ and intrinsic values $(\xi,\iota)$ with the standard errors-in-variables formulation,
  \begin{eqnarray*}
    x_i & = & \xi_i + \epsilon_{x,i} \label{eq-xerr} \\
    y_i & = & \iota_i + \epsilon_{y,i} \label{eq-yerr},
  \end{eqnarray*}
where the errors for each measurement $(\epsilon_{x,i},\epsilon_{y,i})$ are drawn from a multivariate normal distribution with known covariance matrix
\begin{equation*}
\Sigma_i = \left(
\begin{array}{cc}
\sigma_{y,i}^2 & \sigma_{xy,i}\\
\sigma_{xy,i} & \sigma_{x,i}^2
\end{array}
\right).
\end{equation*}
Finally, we assume that the distribution of the intrinsic variable $\xi$ has the probability distribution of a $G$ component univariate normal mixture model, with mixing components $\eta_g$, means $\mu_g$, and standard deviations $\tau_g$. This scenario can be described by the following hierarchical model
\begin{eqnarray}
\label{hm1.eqn} \xi_i | \eta, \mu, \tau^2 &\sim& \sum_{g=1}^G \eta_g \mathcal{N}(\mu_g,\tau_g^2)\\
\label{hm2.eqn} \iota_i | \xi_i, \alpha, \beta, \sigma^2 &\sim& \mathcal{N}(\alpha+\beta\xi_i,\sigma^2)\\
\label{hm3.eqn} y_i, x_i | \iota_i, \xi_i &\sim& \mathcal{N}( (\iota_i \,\, \xi_i)^\top,\Sigma_i).
\end{eqnarray}
The observed-data likelihood can be found by integrating the complete-data likelihood over the missing data $(\xi,\iota)$. Thus, we are able to obtain an equation for the observed-data likelihood in terms of observed quantities, ${\bf x}=(x_1, \ldots,x_n)$, ${\bf y}=(y_1,\ldots,y_n)$, the known covariance matrices
 $\Sigma=(\Sigma_1, \ldots, \Sigma_n)$, and the model parameters, which we write as $\psi = (\eta,\mu,\tau^2)$ and $\theta=(\alpha,\beta,\sigma^2)$:
\begin{eqnarray*}
    p({\bf x},{\bf y}|\theta,\psi) & = & \prod_{i=1}^{n} \iint p(x_i,y_i,\xi_i,\iota_i|\theta, \psi) d\xi_i\,d\iota_i \label{cl1.eqn} \\
         & = & \prod_{i=1}^{n} \iint p(x_i,y_i|\xi_i, \iota_i) p(\iota_i|\xi_i, \theta)   p(\xi_i|\psi)d\xi_i\,d\iota_i \label{cl2.eqn}.
\end{eqnarray*}
This separation into components is possible due to the hierarchical nature of the model. Substituting the probability distributions from Equations~(\ref{hm1.eqn})--(\ref{hm3.eqn}), and integrating over $\xi$ and $\iota$ yields,
\begin{equation}
    p({\bf x},{\bf y}|\theta,\psi)  =  \prod_{i=1}^{n} \sum_{g=1}^{G} \frac{\eta_g}{2\pi |V_{g,i}|^{1/2}}
    \exp \left\{ -\frac{1}{2} ({  w}_i - \zeta_g)^T V_{g,i}^{-1} ({ w}_i - \zeta_g)
      \right\} \label{cl3.eqn}
\end{equation}
where
\begin{eqnarray*}
&w_i = \left( \begin{array}{c} y_i  \\ x_i \end{array} \right) , \quad
   \zeta_g  =  \left( \begin{array}{c} \alpha + \beta \mu_g \\  \mu_g\end{array} \right)  , \label{eq-zeta} \\[1mm]
 &   V_{g,i}  =  \left( \begin{array}{cc}
      \beta^2 \tau_g^2 + \sigma^2 + \sigma^2_{y,i} & \beta \tau_g^2 + \sigma_{xy,i} \\
      \beta \tau_g^2 + \sigma_{xy,i} & \tau_g^2 + \sigma^2_{x,i} \end{array} \right).  &
\end{eqnarray*}
Equation~(\ref{cl3.eqn}) expresses the likelihood of the hierarchal model in terms of the observational data, the mixture model parameters, and the regression model parameters.

\begin{figure}[t!]
\centering
\includegraphics[width=0.60\textwidth]{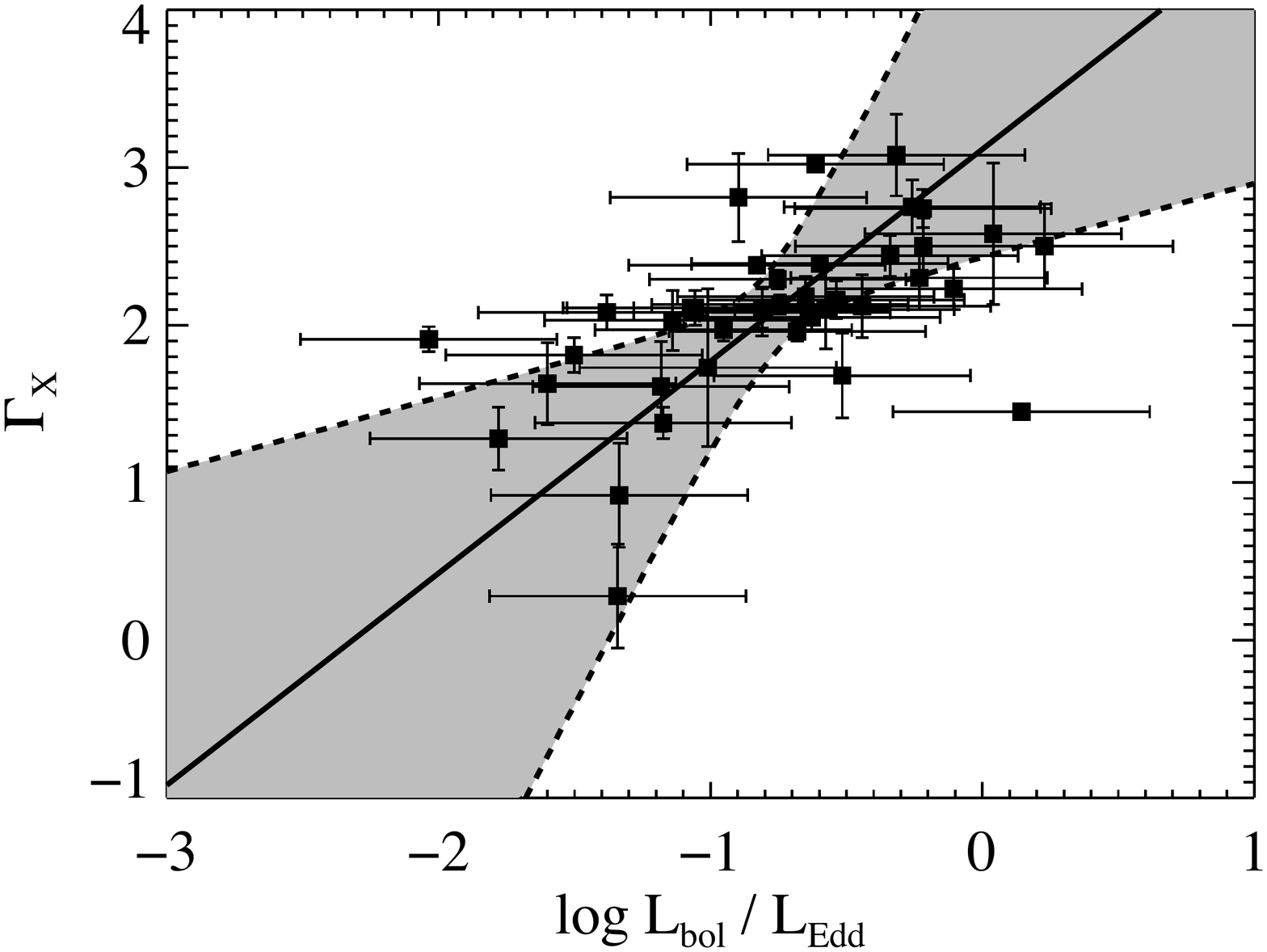}
\caption{ Regression of $\log L_\mathrm{bol}/L_\mathrm{Edd}$ as a function of X-ray photon flux $\Gamma_x$ fit using the hierarchical model using Bayesian inference. The individual uncertainties on the dependent and independent variables are indicated by the error bars (marking one standard deviation). Spread in the observed $\Gamma_x$ is dominated by measurement error rather than intrinsic spread.   \citep{2007ApJ...665.1489K}  \label{quasars.fig}}
\end{figure}

\citet{2007ApJ...665.1489K} provide an example from \citet{2007ApJS..168....1K} where this method is useful for regression analysis. The data consist of 39 quasars that had been observed with the {\it Chandra} X-ray Observatory and SDSS.  Figure~\ref{quasars.fig} is a scatter plot showing two quantities obtained from these data, $\Gamma_x$, the ``X-ray photon index'' -- a measure of the distribution of energies of X-ray photons from the quasars, and $\log L_\mathrm{bol}/L_\mathrm{Edd}$, an estimate of the rate of inflow of matter onto the central black hole in these quasars in terms of the theoretical maximum rate. In this example, the uncertainties on both estimates are independent. Using Bayesian methods with Markov Chain Monte Carlo sampling, a reasonable set of priors, and $G=2$ mixture components, \citet{2007ApJ...665.1489K} estimate $\hat{\alpha} = 3.12 \pm 0.41$, $\hat{\beta} = 1.35 \pm 0.54$, and $\hat{\sigma} = 0.26 \pm 0.11$. From the mixture model parameters, it turns out that the scatter in the independent variable, $\Gamma_x$, is dominated by measurement error.

Implementation of the {\it linmix\_err} algorithm in IDL is available from the IDL Astronomy User's Library\footnote{https://idlastro.gsfc.nasa.gov} and in python  from GitHub.\footnote{https://github.com/jmeyers314/linmix}

\subsection{Deconvolution of distributions from data with heteroscedastic errors and missing information}

The remarkable ability of normal mixture models to closely model many different probability density functions is also used by \citet{2011AnApS...5.1657B} in their Extreme Deconvolution algorithm. This algorithm is used to recover underlying distribution functions in cases where individual measurements have known heteroscedastic uncertainties and/or some variables are missing due to projection effects. For example, for astronomers studying the motions of stars by measuring their position in images of the sky (proper motion), the velocity of a star in three-dimensional space would be projected onto the two-dimensional plane of the sky. In this case, the component of velocity parallel to the line of sight is the missing data.

The model created by \citet{2011AnApS...5.1657B} describes the intrinsic distribution of quantities subject to these observational limitations, with the assumption that the intrinsic distribution is a mixture of multivariate normal distributions. The method addresses the common challenge of ``deconvolution'' in astronomy, but, unlike the conventional deconvolution methods used by astronomers, it allows each measurement to have individual uncertainties.

In this model, the observed data $\mathbf{w}$ are related to the intrinsic values $\mathbf{v}$ through the addition of measurement error $\epsilon_i$. Here, the use of boldface notation indicates that variables $\mathbf{w}$ and $\mathbf{v}$ may be elements of multi-dimensional vector spaces. We assume that the measurement error for each observation $\mathbf{w}_i$ is drawn from a multivariate normal distribution with a known covariate matrix $\mathbf{S}_i$. In addition to uncertainty, \citet{2011AnApS...5.1657B} also include transformation and projection of the data with a projection matrix $\mathbf{R}_i$ to account for missing data.

When including all these effects, the observed data are related to the intrinsic values by the equation,
\begin{equation*}
\mathbf{w}_i = \mathbf{R}_i\mathbf{v}_i+\epsilon_i.
\end{equation*}
The next assumption is that the intrinsic variable $\mathbf{v}$ has the probability distribution of a $G$-component normal mixture model,
\begin{equation*}
\mathbf{v} | \eta, \mathbf{m}, \mathbf{V} \sim \sum_{g=1}^G \eta_g \mathcal{N}(\mathbf{m}_g,\mathbf{V}_g),
\end{equation*}
where $\eta_g$ are the mixing coefficients, $\mathbf{m}_g$ are the means of the components, and $\mathbf{V}_g$ are their covariance matrices. The estimation of these model parameters, denoted as $\theta$, is the desired output of the algorithm.

The observed likelihood equation for this model is then
\begin{eqnarray*}
p(\mathbf{w}_i |\mathbf{R}_i ,\mathbf{S}_i,\theta) &=&\sum_{g=1}^G\int
_\mathbf{v}\mathrm{d}\mathbf{v}\, p(\mathbf{w}_i ,\mathbf{v},g|\theta)\nonumber\\
&=& \sum_{g=1}^G\int_\mathbf{v}\mathrm{d}\mathbf{v}\, p(\mathbf{w}_i |\mathbf{v})
p(\mathbf{v}|g,\theta)p(g|\theta),\nonumber
\end{eqnarray*}
for which,
\begin{eqnarray*}
p(\mathbf{w}_i |\mathbf{v}) &=& \mathcal{N}(\mathbf{w}_i |\mathbf{R}_i \mathbf{v},\mathbf{S}_i),\nonumber\\
p(\mathbf{v}|g,\theta) &=& \mathcal{N}(\mathbf{v}|\mathbf{m}_g,\mathbf{V}_g),\\
p(g|\theta) &=& \eta_g.\nonumber
\end{eqnarray*}
Performing the integral over $\mathbf{v}$ gives a likelihood in the form of a normal mixture model, which is written
\begin{equation*}
p(\mathbf{w}_i |\theta) = \sum_{g=1}^G \eta_g\mathcal{N}(\mathbf{w}_i |\mathbf{R}_i \mathbf{m}_g,\mathbf{T}_{ig}),
\end{equation*}
where the covariance matrices are
\begin{equation*}
\mathbf{T}_{ig}= \mathbf{R}_i \mathbf{V}_g\mathbf{R}_i \T+ \mathbf{S}_i .
\end{equation*}
The log-likelihood for $\theta$, given all $n$ observations $\mathbf{w}_i$, is then
\begin{equation}\label{bovy.eqn}
\ell_o ( \theta) =p(\mathbf{w}_1,...,\mathbf{w}_n| \theta)= \sum_{i=1}^n \log p(\mathbf{w}_i|\theta) =  \sum_{i=1}^n \log \sum_{g=1}^G \eta_g\mathcal{N}(\mathbf{w}_i |\mathbf{R}_i \mathbf{m}_g,\mathbf{T}_{ig}).
\end{equation}
This log-likelihood equation can be used for maximum-likelihood estimation or Bayesian inference to estimate the parameters $\eta_g$, $\mathbf{m}_g$, and $\mathbf{V}_g$ of the intrinsic distribution. \citet{2011AnApS...5.1657B} use an EM approach to address this problem.
Implementation of the Extreme Deconvolution algorithm in Python is available from GitHub.\footnote{https://github.com/jobovy/extreme-deconvolution}

\begin{figure}[t!]
\centering
\includegraphics[width=0.60\textwidth]{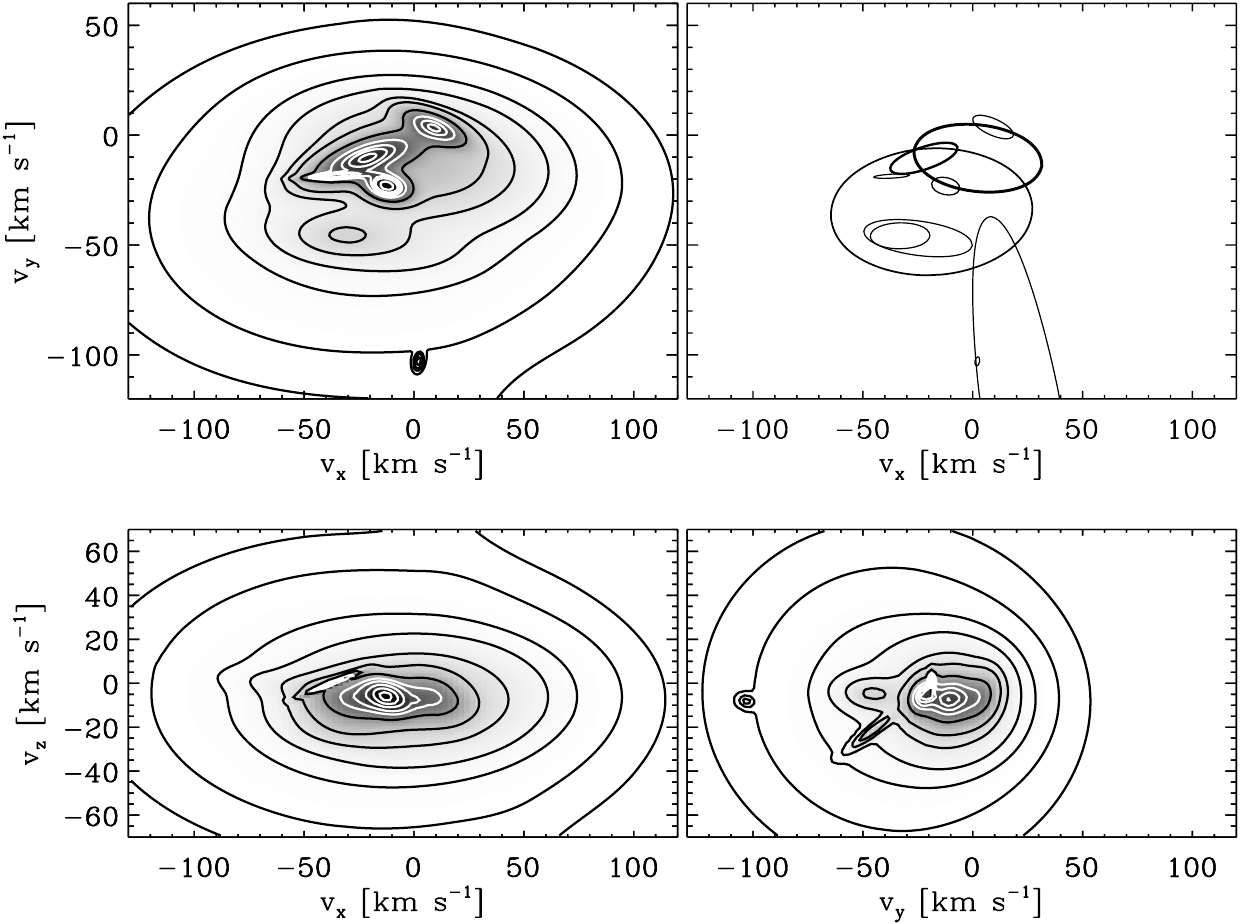}
\caption{The recovered three-dimensional distribution of stellar velocities for {\it Hipparcos} stars in the solar neighborhood is projected onto the $x$--$y$ plane. Density is indicated by the grayscale and contour lines. Clumps in the image are moving groups. \citep{2011AnApS...5.1657B} \label{bovy.fig}}
\end{figure}

The Extreme Deconvolution method has been useful in assessing the structure and kinematics (star motions) of our Milky Way Galaxy.  \citet{2009ApJ...700.1794B} used the algorithm to model the three-dimensional kinematics of $10^4$ stars in the solar neighborhood based on their two-dimensional tangential velocities as measured in the plane of the sky by the {\it Hipparcos} satellite. In this example, the radial velocities of individual stars is unknown and the uncertainties on the tangential velocities depend on the stars' brightnesses and distances. Figure~\ref{bovy.fig} shows the three-dimensional velocity distributions recovered for the  {\it Hipparcos} stars, projected onto the $x$--$y$ plane. Contours and gray-scale show density of sources, with the clumps in the distribution being likely kinematic moving groups. In this example, the Extreme Deconvolution algorithm was successful at identifying major known moving groups based on three dimensional kinematics.

\section{Conclusions}

We have reviewed here both traditional normal mixture models and some unusual types of mixture modeling in astronomical research.  Several lines of reasons suggest there is considerable potential for the expansion of such statistical applications in the future:
\begin{enumerate}

\item The appearance of the phrase ``mixture model'' in astronomical research papers has increased considerably in recent years from a negligible level prior to 2000, to a dozen studies annually around 2010 and several dozen studies around 2015.  This is mostly due to promulgation of a few astrostatistical papers in certain subfields (Ashman et al. 1994; Muratov \& Gnedin 2010; Kelly 2007; Bovy et al.\ 2011).  Knowledge of the broader methodology among astronomers is weak; out of 20,000 papers published annually  barely  two  refer to the authoritative monograph by McLachlan \& Peel (2000). But the actual use is undoubtedly much greater, as most astronomers are not familiar with the label ``mixture models'' and simply view them as a class of multi-component regression models.  The occurrence of mixtures $-$ overlapping populations in a survey, groups of stars or galaxies from different classes, clouds of gas with different velocities in a single structure, and so forth $-$ is very common in astronomy.

\item Astronomers have often grouped points in a $p$-dimensional space $-$ where the dimensions can represent either spatial location or values of some space of observed properties $-$ using heuristic methods.  A common procedure is to construct a decision tree classifier by visual examination without algorithm.  Long-standing classifications of galaxy morphology, active galactic nuclei, supernovae and other classes of cosmic objects are based  on these subjective procedures.  In cases where the property distributions can be represented by multivariate normals or similar distributions, mixture modeling can improve definitions of classes and allow new objects to be associated with these classes in an objective fashion. For example, \citet{2017MNRAS.472.2808D} have refined a traditional heuristic 3-cluster division of galaxy by emission line properties (the Baldwin-Phillips-Terlevich diagram) into a 4-cluster structure using Gaussian mixture models.

\item In cases where astronomers have used objective clustering procedures such as the single-linkage ``friends-of-friends'' algorithm, the resulting classifications are often unstable to arbitrary choices of nonparametric procedures.  Shifting to maximum likelihood estimation of parametric mixture models can lead to more stable and reproducible classifications.  In particular, the number of components emerges from quantitative model selection measures such as the BIC rather than from subjective decisions.

\item Today in the early 21$^{st}$ century, enormous resources are being devoted to wide-field sky surveys in many spectral bands.  Like commercial fishing trawlers that draw a wide variety of sea creatures in a single haul, these surveys collect vast numbers of stars, galaxies, active galactic nuclei, and transient phenomena.  Mixture models should be promoted as important classification tools to treat these problems arising from SDSS, LSST and similar large surveys. These models may be used in advanced statistical modeling of big data, such as the methods by  \citet{2017JCAP...10..036R} for obtaining cosmological parameters from LSST.

\end{enumerate}

Statisticians can help, not only improve interpretation of astronomical data, but can find a rich world of datasets in astronomy for testing methodological developments in mixture modeling.  Nearly the entire research literature is publicly available online in full text through the NASA/SAO Astrophysics Data System (http://adswww.harvard.edu/abstract\_service.html).  A considerable fraction of the data underlying these studies are publicly available.  Petabytes of calibrated images, spectra, and time series are provided by archive centers for both space-based and ground-based observatories.  But derived tabular material may be more useful for the statistician.  These include large billion-object catalogs from surveys such the SDSS, and smaller catalogs and tables from specialized studies.  The Vizier Web service (http://vizier.u-strasbg.fr) gives access to thousands of such tables, including a dozen with more than $10^8$ objects.  Catalogs of cosmic populations with specified properties can be obtained from the NASA/IPAC Extragalactic Database, SIMBAD database, and the International Virtual Observatory Alliance.  However, due to the complexities of the scientific  questions and the individual peculiarities of each survey, we recommend that statisticians exercise their expertise in collaboration with astronomers to address important astronomical and astrophysical questions in a reliable fashion.

\section*{Acknowledgements}

Michael A. Kuhn  is grateful to the Chilean Millennium Institute for Astrophysics (MAS) and CONICET agency for a post-doctoral fellowship, and to the Department of Astronomy at the University of Valpara\'iso for hospitality.  Eric D. Feigelson  is a professor in the Departments of Astronomy \& Astrophysics and of Statistics at Penn State affiliated with the University's Center for Astrostatistics.  His research in astrostatistics is supported by NSF grant AST-1614690 and travel for this work was supported by the MAS.


\bibliographystyle{authordate1}
\bibliography{ch19.bbl}

\end{document}